\documentclass[conference]{IEEEtran}
\usepackage{float}
\usepackage{graphicx}
\usepackage{caption}
\usepackage{subcaption}
\usepackage{amsmath}
\usepackage{array}
\usepackage{tabularx}
\usepackage{multirow}
\usepackage{lipsum}
\usepackage{booktabs}
\usepackage{makecell}

\ifCLASSINFOpdf
\else
\fi
\begin{document}
%
\title{High-Impedance Non-Linear Fault Detection via Eigenvalue Analysis with low PMU Sampling Rates}
%
%
%

\author{\IEEEauthorblockN{Gian Paramo}
\IEEEauthorblockA{\textit{Electrical and Computer Engineering} \\
\textit{University of Florida}\\
Gainesville, FL, USA \\
gparamo@ufl.edu}
\and
\IEEEauthorblockN{Arturo Bretas}
\IEEEauthorblockA{\textit{Distributed Systems Group} \\
\textit{Pacifc Northwest National Laboratory}\\
Richland, WA, USA \\
arturo.bretas@pnnl.gov}
\and
\IEEEauthorblockN{Sean Meyn}
\IEEEauthorblockA{\textit{Electrical and Computer Engineering} \\
\textit{University of Florida}\\
Gainesville, FL, USA \\
meyn@ece.ufl.edu}
\thanks{Gian Paramo is with the Department
of Electrical and Computer Engineering, University of Florida, Gainesville,
FL, USA e-mail: gparamo@ufl.edu.}}

\maketitle

\begin{abstract}
This work presents a hybrid data-driven and physics-based framework for high-impedance fault detection in power systems. An innovative method based on eigenvalue analysis is expanded and validated. Phasor Measurement Unit data is used to estimate eigenvalues corresponding to the powerlines being monitored. Eigenvalue statistics are then tracked and evaluated. 
Faults are detected as they drive eigenvalues outside of their normal zones. 
This technique holds several advantages over contemporary techniques: It utilizes technology that is already deployed in the field, it offers a significant degree of generality, and so far it has displayed a very high-level of sensitivity without sacrificing accuracy. Validation is performed in the form of simulations based in the IEEE 13 Node System and non-linear fault models. Test results are encouraging, indicating potential for real-life applications. 
\end{abstract}


\IEEEpeerreviewmaketitle

\section{Introduction}
\IEEEPARstart{T}{he} non-linear relationship between currents and voltages, the presence of electrical arcs, and most importantly, the low current magnitudes make it difficult for traditional overcurrent protection to detect high-impedance faults (HIFs). This gap in protection creates exposures in terms of service reliability, equipment integrity, and safety. These concerns are far from hypothetical as numerous fatalities have been attributed to HIFs \cite{re1, re2}. 
Advances in technology, in particular, the introduction of the Phasor Measurement Unit (PMU), offer new tools to overcome this challenge. By leveraging the high sampling rates of PMUs combined with their ability to synchronize measurements, input-output relationships can be established at the ends of a powerline. This makes it possible to conduct an in-dept analysis of the behavior and trends of the system. 
Presently, modern digital relays are capable of providing PMU measurements, however, their sampling rate is only 30 Hz \cite{sel421}. This is the equipment demographic this method aims to leverage. Instead of waiting years, or possibly decades for the hypothetical scenario where high sampling rate PMUs are universally available, this work aims to contribute towards the solution of the HIF problem considering technology that is currently available and already in service. 

\section{Related Work}

\begin{figure}
\centering
\captionsetup{justification=centering}
\includegraphics[width=8.5cm]{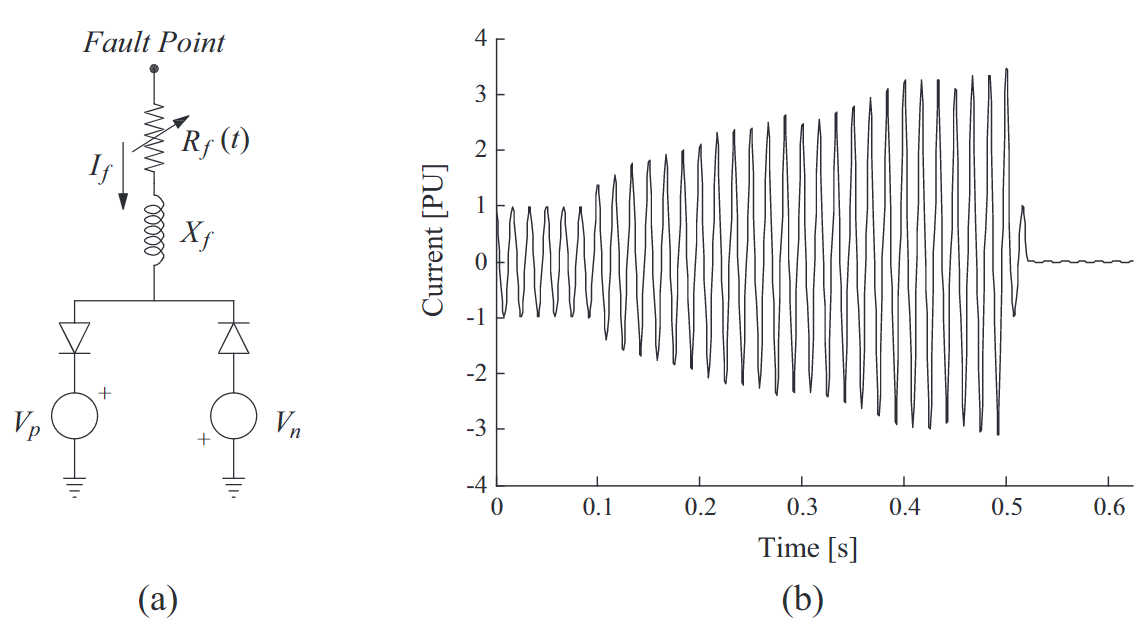}
\caption{Anti-parallel source-diode HIF model \cite{FERRAZ}.\label{HIF}}
\end{figure}

The late 80's and early 90's saw a new trend in HIF detection where distortions in the waveforms caused by HIFs were used to characterize the event \cite{BENNER, AUCOIN}. Mathematical tools such as the Fourier transform were used for this purpose, as the fundamental component of the waveform was separated from the distortion generated by the non-linear elements of HIFs \cite{SULTAN, YU}. During this period, HIFs models were developed, and some researchers even had the foresight to use machine learning (ML) techniques such as artificial neural networks (ANNs) for the purpose of detecting HIFs \cite{LUB}. This period laid the foundation for many of the contemporary HIF solutions available in literature today.  A very influential HIF model has been the so called anti-parallel source-diode model. This model has become a staple in modern HIF research.  Variants of this model are used in \cite{ref8, JUN, FERRAZ}. Figure \ref{HIF} illustrates a version of the anti-parallel source-diode model, and its corresponding current. 
 
Since its inception this model has proven to be a strong tool in the design and analysis of HIF detection. This is due to its ability to emulate a wide range of non-linear features seen in HIFs, including the harmonics produced by electrical arcs and variable impedance angles. While powerful in the research environment, the implementation or integration of this model into larger systems presents significant challenges. Chiefly among them is the inherit need to run a large number of simulations to extract features that are then used for fault detection, as seen in \cite{JUN}. Despite these challenges, the anti-parallel source-diode model remains a cornerstone of modern HIF research.  

Linear estimators are used in \cite{FERRAZ} for the detection of HIFs. Detection is performed by approximating fault parameters retrieved via sliding time windows. 
The main challenge encountered by these types of solutions is that critical features can only be captured during very specific time windows that last only a few milliseconds.  

A variation of the wavelet transform is used for HIF detection in \cite{costa}. While powerful, one of the main drawbacks of the wavelet transform is that it requires measurements to be taken at high sampling rates and can only be supported by specialized equipment.  

The work presented in \cite{cui} utilizes the anti-parallel source-diode HIF model developed in \cite{ref8} to train a semi-supervised learning algorithm. 
The results produced by this solution were encouraging, in particular in the presence of noise and in terms of accuracy; however, the level of complexity in regards to implementation was increased in this solution. First, the need to run a large number of simulations using the model in \cite{ref8} remains. Second, implementing a learning algorithm could require specialized skills and significant engineering time. 


In \cite{choi}, an adaptive and settingless protection scheme based on PMUs is presented. This is a decentralized scheme that assumes key components and locations are equipped with PMUs. 
While promising, this method has some limitations; in particular that some parameters have to be calculated manually before implementation. 

A technique developed for anomaly detection in dynamic state estimation is presented in \cite{ref4}.  The algorithm consists of several detector types that are used in unison. 
Being a data-driven solution, generality is one this solution's highlights. 
This solution delivered encouraging results, but as presented in \cite{ref4}, the framework was not a complete solution and only provided data analysis for results produced by other estimators.  

Fault location in active distribution networks is addressed in \cite{ref5}. 
A novel aspect of this technique is its use of supervisory information to adjust estimation parameters autonomously and in real-time. Some possible drawbacks of the technique include the possible information bottlenecks that could be created if a fault were to impact multiple lines in a very large system. Also, details of how and how much supervisory information is required at each estimator is not clear.

A scheme for back-up protection based on PMUs was developed in \cite{ref6}. 
The exact location of a disturbance is estimated via weighted least squares. 
The results produced by this solution were encouraging. 
That said, the solution appears to be limited to system topologies of moderate dimension. 

 $\mu$-PMUs are leveraged in \cite{ref9} to detect disturbances in distribution feeders. 
 This technique is able to detect HIFs of resistance values on the lower end of the HIF spectrum (higher current magnitudes). 
 An extension of \cite{ref9} was presented in \cite{ref9.5}, where ML techniques were used to extract additional features from data reported by PMUs. This extension managed to deliver performance that was more robust compared to the original technique; however, the use of ML techniques, as previously discussed, can bring challenges in terms of implementation and replication. 

\section{Summary of Contributions}

The framework presented in this work is based on online change detection, and it is aimed at finding a compromise between model based solutions and data-driven 'blackbox' techniques. This method operates as a setting-less protection scheme, backed by a simple physics based model that is estimated from online data. 
The end product is a solution that doesn't waste resources trying to build and analyze a database for an event of stochastic nature, and doesn't require specialized skills sometimes required by AI applications. 

The original formulation of this work in \cite{OG}, produced encouraging results; however, several key aspects required further consideration. For instance, faults of a non-linear and time-varying nature were not evaluated. High sampling frequencies (120 Hz) were used, and the overall system was a relatively simple one. In this work, the solution is tested in the presence of time-varying fault resistances, harmonics, and other non-linear elements related to HIFs. Testing carried out as part of this work shows that the solution is capable of detecting HIFs, even with PMU sampling rates as low as 30 Hz. This gives the framework a significant advantage in terms of practicality compared to other techniques, where high sampling rates, in some cases, over 120 Hz are used for parameter identification \cite{GP}. 

In another step in addressing the technological limitations of the grid, this solution is intended to operate as a de-centralized solution to ease the burden on communication networks \cite{GP, Bose, Naduv}.
Finally a mathematical model that supports the principles and theory behind this work is presented in Section IV, equations 1 through 6. 
\section{Expanded Eigenvalue HIF Detection}
\begin{figure}[H]
\centering
\captionsetup{justification=centering}
\includegraphics[width=4.5cm]{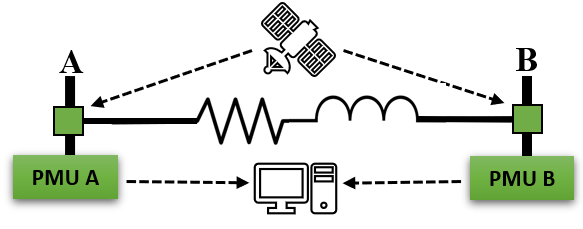}
\caption{Simplified PMU System.\label{pmusat2}}
\end{figure}

\begin{figure}[H] 
\centering
\captionsetup{justification=centering}
\includegraphics[width=7cm]{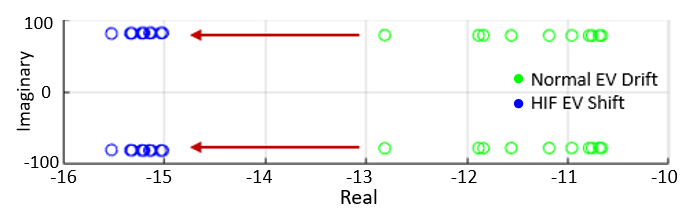}
\caption{Eigenvalue Drift Normal and HIF Conditions.\label{EV Drift}}
\end{figure}

\begin{figure}[H] 
\centering
\captionsetup{justification=centering}
\includegraphics[width=7cm]{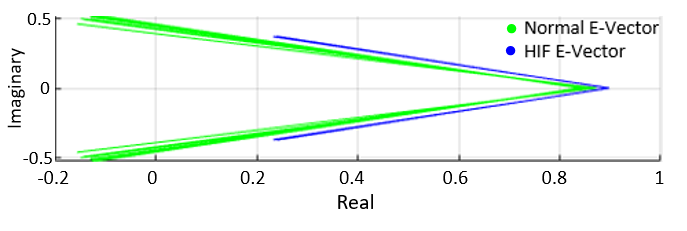}
\caption{Eigenvectors Under Normal and HIF Conditions.\label{E Vector}}
\end{figure}

\begin{figure}[H]
\centering
\captionsetup{justification=centering}
\includegraphics[width=7cm]{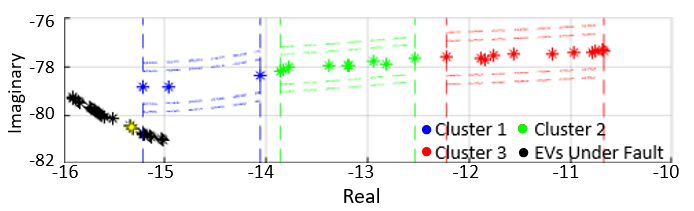}
\caption{Protection Zones and HIF Drift.\label{675 Fault}}
\end{figure}

\begin{figure}[H] 
\centering
\captionsetup{justification=centering}
\includegraphics[width=4cm]{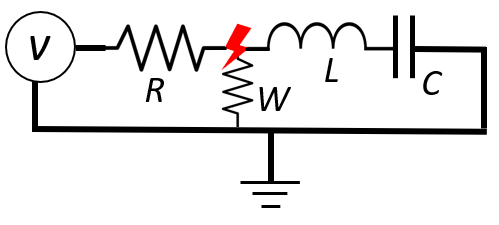}
\caption{RLC Circuit Under Fault.\label{SS2}}
\end{figure}

Prior work in \cite{OG} begins by realizing a set of eigenvalues from either historical data or from real-time data. The estimated physical model, and related eigenvalues are derived from an apparent impedance calculated from voltage and current readings are taken at the ends of the power line as illustrated by Figure \ref{pmusat2}. 
Each powerline, due to its unique impedance and the characteristics of the load, has a distinctive projection in the eigenvalue space. This projection is dynamic and changes over time, reflecting the operating conditions of the system. In highly dynamic systems, the eigenvalue projection can span across a large area in the eigenvalue space, leading to significant changes in eigenvalue location. 
While these changes can sometimes seem dramatic, they pale in comparison to the changes produced by disturbances such as faults, where the eigenvalue projection is dominated by the characteristics of the fault. Although HIFs don't usually produce shifts in position as drastic as those seen in low impedance faults, HIFs still manage to drive eigenvalues away from their normal zones. This sensitivity is what this technique aims to exploit for HIF detection. Figure \ref{EV Drift} depicts changes in the drift of eigenvalues due to a fault with a current magnitude of less than $10\%$ of nominal. Figure \ref{E Vector} illustrates the behavior of the eigenvectors under normal and fault conditions. 
The concept of zones of normal operation is derived from the principles utilized in Distance and Out-of-Step (OOS) relaying. In these schemes, impedance is mapped onto a complex plane and bounded inside zones of protection.  
Polynomial curve fitting is used to define the boundaries of the zones of normal operation. 
These zones adapt as new data arrives. The update intervals and the number of eigenvalues projected are defined by the user. 
In order to facilitate the creation of the zones of protection and to also increase the selectivity of the protection scheme, eigenvalue locations are broken into clusters. 
Figure \ref{675 Fault} displays the zones of protection generated by the algorithm along with the shifts in position as HIFs are introduced.
This framework utilizes real-time data to estimate a simple physics-based model. This model is then used for change detection. Mathematically, during normal conditions the voltage-current relationships of the powerline can be described with a simple differential equation:
\[ v(t) = Ri(t) + L\frac{di(t)}{dt} + v_{c}(t)  \tag{1} \]
which corresponds to the circuit in Figure \ref{SS1}:

 \begin{figure}[H] 
\centering
\captionsetup{justification=centering}
\includegraphics[width=4cm]{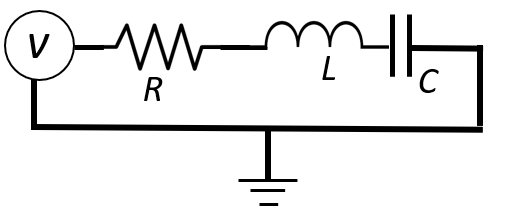}
\caption{RLC Circuit. \label{SS1}}
\end{figure}
Where $v(t)$ is the voltage at the source, $R$ is the resistance of the line, $L$ is the inductance, $C$ represents the capacitance supplied by the capacitor, and $v_{c}(t)$ is the voltage  across the capacitor. Taking this system into state space produces the following representation:
\[ \Dot{X} = \begin{bmatrix} \tag{2}
-\frac{R}{L}&-\frac{1}{L}\\
\frac{1}{C}&0\\
\end{bmatrix} 
\begin{bmatrix}
i(t)\\
v_{c}(t)\end{bmatrix}
+
\begin{bmatrix}
\frac{1}{L}\\
0\end{bmatrix}
\begin{bmatrix}
v(t)\\
\end{bmatrix}
\] 
For this simple model, two state variables are used. These are the current $i(t)$, and the voltage at the capacitor $v_{c}(t)$ . When the system is subjected to a fault with a constant impedance, as illustrated by Figure \ref{SS2}, the state space representation becomes that of an $RLC$ circuit with a parallel branch to ground. 

Using the same $RLC$ circuit but now accounting for a connection to ground between $R$ and $L$, the state space representation becomes: 
\[ \Dot{X} = \begin{bmatrix} \tag{3}
-\frac{RW}{L(R+W)}&-\frac{1}{L}\\
\frac{1}{C}&0\\
\end{bmatrix} 
\begin{bmatrix}
i_{o}(t)\\
v_{c}(t)\end{bmatrix}
+
\begin{bmatrix}
\frac{W}{L(R+W)}\\
0\end{bmatrix}
\begin{bmatrix}
v(t)\\
\end{bmatrix}
\] 

In this case $W$ represents a constant fault impedance. It is evident that the introduction of the fault will cause a change in the eigenvalues. When the system is subjected to a fault with the profile described in \cite{ref8}, the term $W$ becomes $w$. $w$ can be defined as \cite{JUN}:
\[ w = R_{p}(\tau)i_{p}(t) + v_{p}\textit{sgp}[i(t)] + R_{n}(\tau)i_{n}(t) +  v_{n}\textit{sgn}[i(t)]    \tag{4} \]

  \begin{equation}
  \textit{ sgp}[i(t)]=
    \begin{cases}
      1, & \text{if}\ i(t)>0 \\
      0, & \text{if}\ i(t)\leq 0 \tag{5} 
    \end{cases}
  \end{equation}  \begin{equation}
  \textit{ sgn}[i(t)]=
    \begin{cases}
      0, & \text{if}\ i(t)>0 \\
      -1, & \text{if}\ i(t)\leq 0 \tag{6} 
    \end{cases}
  \end{equation}

  Where $\textit{sgp}[i(t)]$ and $\textit{sgn}[i(t)]$ represent the harmonic components associated with electrical arcs. Resistance $R(\tau)$ is a Gaussian random process with upper and lower limits and an update interval $\tau$.
 The presence of noise was considered in \cite{OG}. It was observed that constant noise does not have a significant impact in the performance of the framework. This because the solution assumes that noise is part of the normal behavior of the system and it learns to ignore it. 


\section{Case Studies}

Validation was performed via simulations in the IEEE 13 Node System. Loads were given a dynamic profile, and follow the trends seen during a typical October day in Houston, TX \cite{ref12}. Three locations were faulted: Fault 671 corresponds to a fault between bus 632 and bus 671. Fault 675 takes place between buses 692 and 675. Fault 634 takes place behind bus 634 (between the bus and its load). The faults were modeled per \cite{ref8}, with magnitudes ranging from 6.7\% to 13.3\% of the nominal line currents. The faults also include harmonic distortion and resistance variations consistent with \cite{ref8}. Fault values are listed in Table \ref{tab1}. 

\begin{table}[H]
\caption{Fault Magnitudes}\label{tab1}
\begin{center}
\begin{tabular}{ccccccc} 
\toprule
\thead{Current [A]} & \thead{Fault 671} & \thead{Fault 675} & \thead{Fault 634}\\ 
\midrule
Nominal & 207 & 98 & 190 \\
Fault DC Component & 1.98 & 2.07 & 3.77 \\
Fault Second Harmonic & 0.89 & 1.02 & 1.69 \\
Fault Third Harmonic & 2.17 & 2.19 & 3.45\\
Fault RMS & 14 & 13 & 17 \\
\hline
\end{tabular}
\end{center}
\end{table}

\begin{figure}
\centering
\captionsetup{justification=centering}
\includegraphics[width=8.5cm]{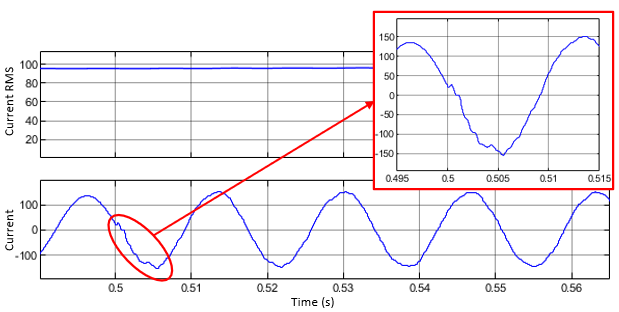}
\caption{Current Waveform at the Relay.\label{wave}}
\end{figure}

Each location was faulted twenty four times, the equivalent of one fault every hour of the day. This is done to examine the effectiveness of the technique under dynamic loading conditions.
During fault conditions, the behavior of the waveforms is similar to those seen in \cite{ref8}, which is expected. Figure \ref{wave} corresponds to the current seen by relays for a fault at location 675. 
The typical features many HIF detection techniques extract and then use to identify these faults can be appreciated. Circled is the initial current spike seen at the start of the fault. The accompanying harmonics and distortion can also be observed. It must be noted that, the time window when these features can be captured is of only 7.5 ms. These limited opportunities for detection along with the unnecessary training and classification done by mainstream HIF detection techniques are what this work eliminates. 

During fault events eigenvalues tend to congregate at one particular location in the eigenvalue space. This is because the position and drift of the eigenvalues during a fault is dominated by the characteristics of the fault, as previously presented. This shift can be appreciated in Figures \ref{EV Drift} and \ref{675 Fault}. The distance between eigenvalues is inversely proportional to the magnitude of the fault current, meaning that as the fault current increases, the eigenvalues move closer together. Metrics produced by this behavior are presented in the results listed in Table \ref{ky}.

\begin{table}[H]
\caption{Eigenvalue Shift Metrics due to HIFs}\label{tab2}

\begin{center}
\begin{tabular}{cccccccc} 
\toprule
\thead{Fault} & \thead{EV Mean} & \thead{EV Standard Div.} \\ 
\midrule

671 (Pre-Fault) & 11\angle-90 & 0.1 \\
671 (Fault) & 376\angle-168 & 119.1 \\
675 (Pre-Fault) & 79\angle-99 & 1.4 \\
675 (Fault) & 82\angle-101 & 0.6 \\
634 (Pre-Fault) & 486\angle-175 & 137.6 \\
634 (Fault) & 11\angle-90 & 0.001 \\

\hline
\end{tabular}
\end{center}
\label{ky}
\end{table}

All fault cases were correctly identified as the eigenvalues drifted outside of their respective zones of protection. The introduction of faults at Locations 671 and 634 produced dramatic shifts in eigenvalue locations. At Location 675, despite seeing relatively minor deviations during fault conditions, the algorithm was still able to identify the disturbance. The results for 675 are illustrated in Figure \ref{675 Fault}.
\subsection{Comparison with Overcurrent Relays}
When information is limited, relays can be set with a basic 20\% margin over the expected nominal current \cite{relaybook}. To highlight the challenge of HIF detection by means of OC protection, a conservative margin of 15\% was used in the following examples.  Table \ref{tab3} presents basic parameters used in OC protection. 
$M$ represents the ratio of the fault current and the relay tap.   
\begin{table}[H]
\caption{OC Relay Parameters}\label{tab3}
\begin{center}
\begin{tabular}{ccccccc} 
\toprule
\thead{Parameter} & \thead{Fault 671} & \thead{Fault 675} & \thead{Fault 634}\\ 
\midrule
Pick-Up & 238 & 112.7 & 218.5 \\
CT Ratio & 250:5 & 150:5 & 250:5 \\
Tap & 4.7 & 3.75 & 4.37 \\
Fault at Relay [A] & 4.42 & 3.7 & 4.14\\
M & $<$ 1 & $<$ 1 & $<$ 1 \\
\hline
\end{tabular}
\end{center}
\end{table}


Despite the higher precision offered by digital relays, in order for the relay to operate, $M$ must be greater than 1, as defined by the equations in \cite{sel751}. Since the values of $M$ remain below 1, the HIFs remain undetected. In these scenarios the framework presented in this paper clearly outperformed OC relays.
\subsection{Performance at Different Sampling Rates}
Finally, the performance of this framework at different PMU sampling rates is examined. For this test, Fault Location 675 is faulted once per hour, over an eight hour period. This test is carried out three times, with each test scenario using a different PMU sampling rate. The remaining system parameters are the same as the previous test. Results are provided in Table IV.    
\begin{table}[H]
\caption{Performance at Varying Sampling Rates}\label{tab4}
\begin{center}
\begin{tabular}{ccccccc} 
\toprule
\thead{Sampling Rate (Hz)} & \thead{30} & \thead{60} & \thead{120}\\ 
\midrule
Mean (Pre-Fault, base of 79) & 1\angle-82 &  2\angle-98 & 4\angle-98\\
Standard Deviation (Pre-Fault) & 0.52 & 1.05 & 2.1 \\
Mean (Fault, base of 82) &  1\angle-100 & 2\angle-100  & 4\angle-100  \\
Standard Deviation (Fault) & 0.7 & 0.3 & 0.7\\

\hline
\end{tabular}
\end{center}
\end{table}

As the sampling rate increases, the eigenvalues appear to be multiplied by a value corresponding to the sampling rate. For instance, the mean at 120 Hz is twice the mean at 60 Hz, and four-times the mean at 30 Hz. Higher sampling rates offer better eigenvalue separation which could lead to more robust results. Despite these differences, the framework was successful in identifying the HIFs at each PMU sampling rate.

\section{Conclusion}
A innovative framework for the detection of HIFs was expanded and validated in this work. Combining elements from well known protection techniques with the high sampling rates of PMUs, a robust HIF detector was presented. This framework offers a high degree of generality, which decreases the complexity of a possible installation. Another highlight of this technique is that it eliminates the need to perform extensive simulations to derive fault parameters, as it is commonly done by solutions based on the anti-parallel source-diode HIF model. 
Testing was conducted in IEEE simulation environments using a popular HIF model~\cite{FERRAZ}. Test results highlight the framework's generality, sensitivity, and accuracy. 

Future work will investigate the application of mathematical tools to alleviate noise related limitations. Fault location will also be addressed in future work. The goal is to develop a formal protection philosophy based on this framework. 

\ifCLASSOPTIONcaptionsoff
  \newpage
\fi


\end{document}